%% file: main.tex
\pdfoutput=1
\documentclass[11pt]{article}

\usepackage{acl}

\usepackage{times}
\usepackage{latexsym}
\usepackage{booktabs}
\usepackage[T1]{fontenc}

\usepackage[utf8]{inputenc}
\usepackage{microtype}

\usepackage{inconsolata}

\usepackage{algorithm}
\usepackage{algpseudocode}
\usepackage{graphicx}
\usepackage{amsmath}
\usepackage{bm}
\usepackage{hyperref}
\usepackage{listings}
\usepackage{xcolor}
\usepackage{caption}
\usepackage{url}

\lstdefinelanguage{json}{
    morestring=[b]",
    morecomment=[l]{//},
    morekeywords={true,false,null},
    sensitive=false,
}

\newcommand{\productname}{OASBuilder}

\title{OASBuilder: Generating OpenAPI Specifications from Online API Documentation with Large Language Models}

\author{
  Koren Lazar\thanks{Corresponding author: \texttt{koren.lazar@ibm.com}} \and
  Matan Vetzler \and
  Kiran Kate \and
  Jason Tsay \and
  David Boaz \\
  \bf{Himanshu Gupta} \and
  Avraham Shinnar \and
  Rohith D Vallam \and
  David Amid \\
  \bf{Esther Goldbraich} \and
  Guy Uziel \and
  Jim Laredo \and
  Ateret Anaby Tavor \\
  IBM Research
}

\begin{document}
\maketitle
\begin{abstract}
AI agents and business automation tools interacting with external web services require standardized, machine-readable information about their APIs in the form of API specifications. However, the information about APIs available online is often presented as unstructured, free-form HTML documentation, requiring external users to spend significant time manually converting it into a structured format.
To address this, we introduce \productname{}, a novel framework that transforms long and diverse API documentation pages into consistent, machine-readable API specifications. This is achieved through a carefully crafted pipeline that integrates large language models and rule-based algorithms which are guided by domain knowledge of the structure of documentation webpages.
Our experiments demonstrate that \productname{} generalizes well across hundreds of APIs, and produces valid OpenAPI specifications that encapsulate most of the information from the original documentation.
\productname{} has been successfully implemented in an enterprise environment, saving thousands of hours of manual effort and making hundreds of complex enterprise APIs accessible as tools for LLMs.

\end{abstract}

\maketitle

\section{Introduction}
\input{oasbuilder/chapters/1-introduction}
\section{OpenAPIs and Documentation Websites}
\label{sec:openapis-and-documentation-websites}
\input{oasbuilder/chapters/2-problem-description}

\section{\productname{}}
\label{sec:OASGen}
\input{oasbuilder/chapters/3-oasbuilder}

\section{Experiments}
\label{sec:experiments}
\input{oasbuilder/chapters/4-results}
\section{Related work}
\label{sec:related-work}
\input{oasbuilder/chapters/5-related-work}

\section{Conclusions}
\label{sec:conclusions}
\input{oasbuilder/chapters/6-conclusions}
\bibliography{custom}

\appendix
\section{Appendix}
\input{oasbuilder/chapters/7-appendix}



\end{document}

%% file: oasbuilder/chapters/1-introduction.tex
AI agents have gained significant popularity in automating tasks across diverse domains, from finance to customer service~\citep{george2023review}, complementing traditional rule-based business automation systems.
Both AI agents and automation systems depend on various external APIs to function effectively. 
For AI agents, APIs serve as tools to access external resources such as real-time data and integration with external services, while for automation systems, APIs are integral to building automated workflows.
However, efficient interaction with these APIs requires that their information be provided in a standardized, machine-readable format.
The OpenAPI Specification (OAS)\footnote{https://swagger.io/specification/} is the leading format for documenting REST APIs~\citep{espinoza2020mapping}, providing a structured, compact representation compatible with large language model (LLM) frameworks such as LangChain~\citep{langchain}.
Unfortunately, many API providers do not provide standardized API specifications.
Our analysis of the 14 most popular APIs on Postman for 2023\footnote{\url{https://www.postman.com/explore/most-popular-apis-this-year}} revealed that only five providers publicly share their OAS (see Appendix~\ref{subsec:appendix-most-popular-urls} for details). Instead, most offer online API documentation presented as HTML webpages with human-readable hypertext describing the API operations. These webpages frequently lack structural consistency and fail to follow standard conventions~\citep{danielsen2013}.
As a result, developers often need to manually convert documentation into OAS format, a labor-intensive error-prone task, especially for real-world APIs, which are typically large and complex. This challenge has sparked the search for automated solutions to convert API documentation webpages into OAS documents~\citep{cao2017automated, Yang-2018, bahrami2020automated, androvcec2023using}. 
However, existing approaches, whether based on automatic parsing or the direct application of LLMs, have consistently struggled to produce accurate and complete OAS documents. These challenges stem from significant variability in API documentation formats, inconsistencies in layout, the presence of embedded JavaScript-generated content, and the considerable length of documentation webpages, often amounting to millions of words.

To address these challenges, we introduce \productname{}, an innovative LLM-based end-to-end framework for automating the generation of OAS from API documentation webpages.
\productname{} employs a multi-stage approach, breaking the OAS generation process into smaller, manageable subtasks. It scrapes API documentation pages, segments them into sections corresponding to individual API operations, and filters out irrelevant content.
Then, the documentation for each API operation (analogous to a function) is then translated into OAS via parallel LLM calls.
Finally, \productname{} provides an intuitive platform for manual validation and editing, supported by evidence from the source webpage and AI-based tools for refining the OAS, ensuring a reliable, high-quality result while significantly reducing manual effort.



To the best of our knowledge, \productname{} is the \textit{first LLM-based automated system} for generating OAS from API documentation pages. 
Our empirical experiments highlight its ability to handle diverse API documentation formats, producing accurate and comprehensive OAS documents. Furthermore, \productname{} has been successfully deployed in an enterprise environment, where it has generated hundreds of API specifications, saving developers thousands of hours of work.\footnote{\url{https://www.ibm.com/docs/en/watsonx/watson-orchestrate/current?topic=skills-using-openapi-builder}.}

%% file: oasbuilder/chapters/2-problem-description.tex
\begin{figure}
    \centering
     \includegraphics[width=.9\columnwidth]{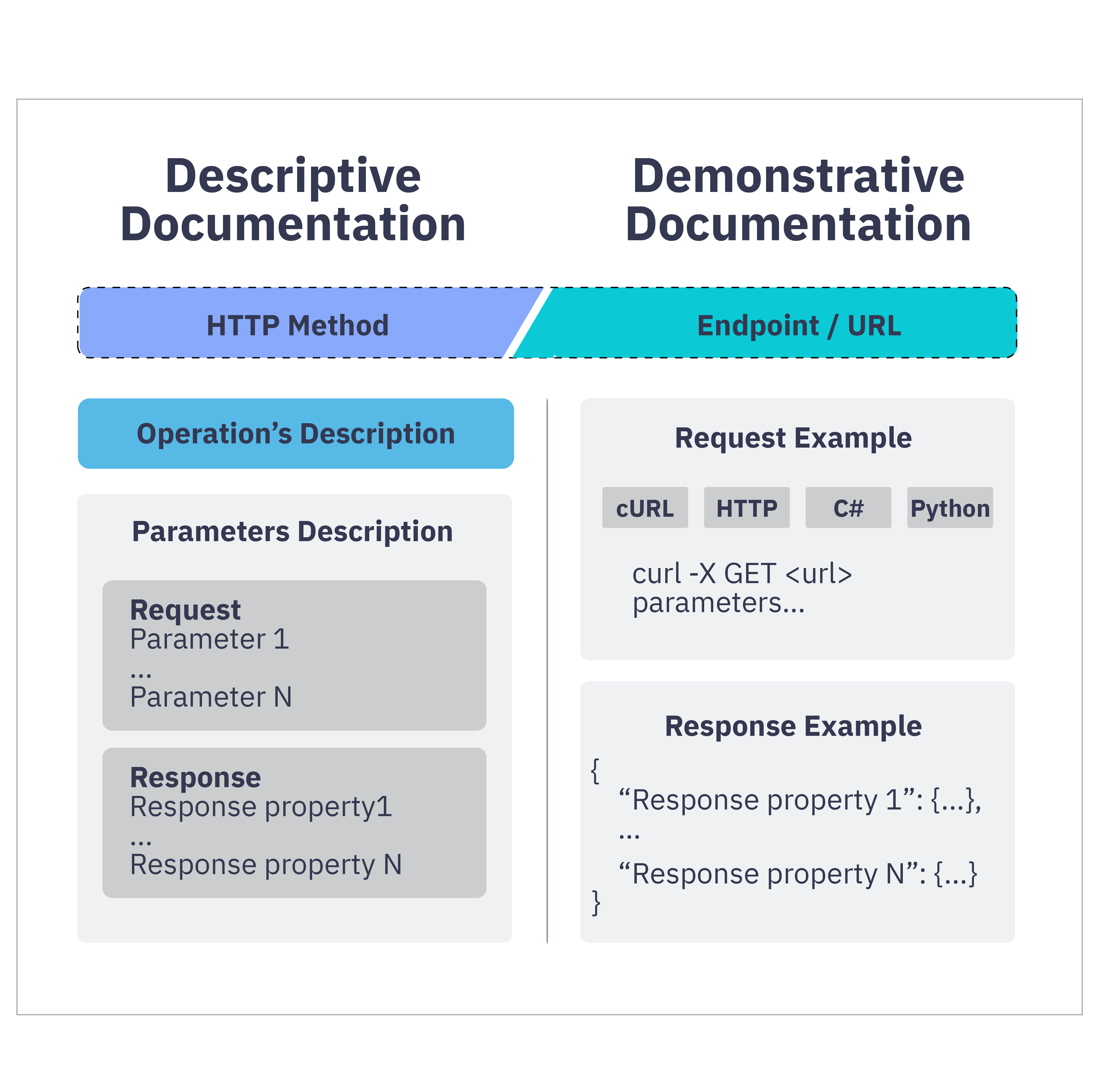}
    \caption{\label{fig:webpage-example} 
    Typical structure of API operation documentation on webpages, featuring the descriptive documentation on the right, the demonstrative documentation on the left, and the operation signature at the top, which can appear on either side.
    A real-world example from Shopify API can be found in Appendix~\ref{subsec:appendix-api-documentation-webpage}.
    }
\end{figure}

The OpenAPI Specification (OAS) is a standardized framework for formally describing RESTful APIs, specifying their operations (analogous to functions or tools), authentication mechanisms, and other operational details. In enterprise applications, OAS documents are typically extensive, comprising numerous operations and deeply nested request and response objects, and frequently span thousands of lines. A minimal example is presented on the right side of Figure~\ref{fig:oasbuilder-system}.

OAS is especially valuable for AI agents as it offers a concise and standardized representation of each API. This allows LLMs to dynamically understand and interact with multiple APIs—a crucial ability for autonomous agents that must reason about and utilize external services.
Moreover, OAS facilitates automation in development processes, such as generating client libraries, server stubs, and documentation, streamlining workflows, reducing manual effort, and minimizing errors.




Despite these benefits, many API service providers only offer online documentation in HTML format, which often lacks consistency in structure or adherence to any standard conventions.
Although this documentation does not adhere to any convention which makes automatic parsing infeasible, document pages often contain recurring semantic components which complements each other.
As shown in Figure~\ref{fig:webpage-example}, these components generally consist of an \textbf{operation signature}, specifying the HTTP method and path (e.g., \texttt{GET /status}); \textbf{descriptive documentation}, providing a textual overview of the operation's purpose, security details, and a tabular breakdown of request and response fields, including field names, data types, formats, required/optional status, and descriptions; and \textbf{demonstrative documentation}, featuring usage examples such as a sample request (e.g., a cURL command) and a sample response (e.g., a JSON object). For a real-world example of such documentation, see Appendix~\ref{subsec:appendix-api-documentation-webpage}.

Although \productname{} leverages each of these components to generate a more comprehensive specification, its sole assumption is the presence of an operation signature or a request example to identify the operations, as detailed in Section~\ref{subsec:scraping}.

%% file: oasbuilder/chapters/3-oasbuilder.tex
\begin{figure*}
    \centering
     \includegraphics[width=\textwidth]
    {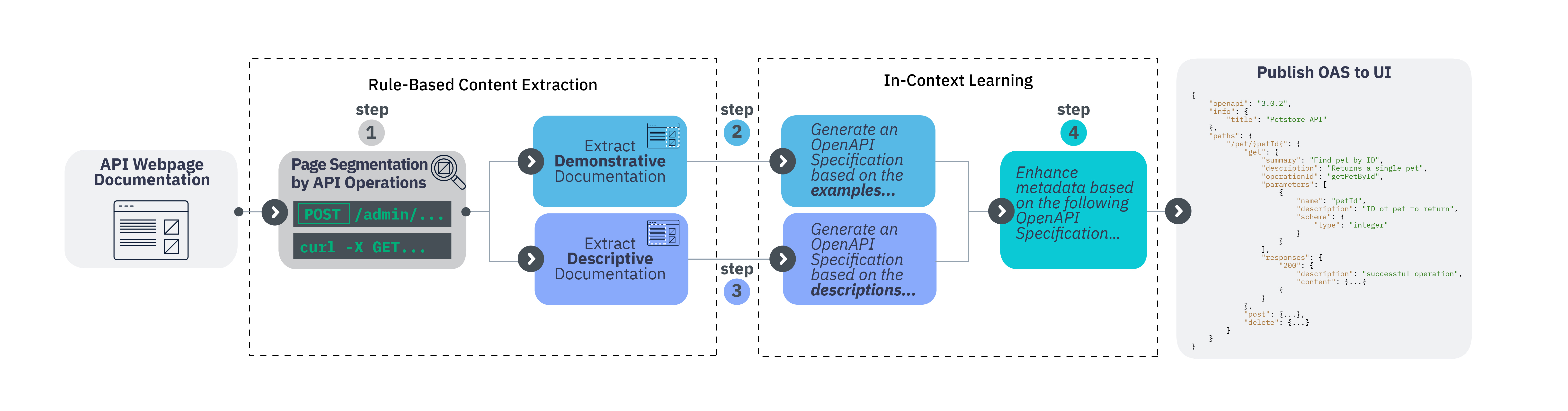}
    \caption{\productname{} pipeline: The system first segments the webpage based on identified API operations. Then, for each operation, it searches for demonstrative and descriptive documentation. The documentations are then processed by an LLM using in-context learning to generate two partial OAS, which are merged into a complete, grounded OAS. The OAS is subsequently enhanced to fill in missing metadata without external resources. Finally, the OAS is presented to the user for final revisions. 
    }
    \label{fig:oasbuilder-system} 
\end{figure*}

\productname{} consists of three main components: (1) an automated method for generating an OAS from a webpage, described in Sections~\ref{subsec:scraping}, \ref{subsec:openapi-spec-generation}, and \ref{subsec:enrichment-specrawler}; (2) AI-powered enhancement of the OAS, detailed in Section~\ref{subsec:enhancement}; and (3) a user-friendly interface for viewing, editing, and validating OAS documents, which also integrates the AI-powered enhancement to further accelerate the final revisions.

Given the potential length of documentation and LLM input limits, the pipeline adopts a modular approach. The generation task is divided into smaller sub-tasks, whose outputs are then combined to create a cohesive result.

Throughout the pipeline, LLM generation was based on in-context learning, as fine-tuning was not feasible due to the lack of labeled data. It is assumed that the LLM was exposed to OAS during its pretraining.
Figure~\ref{fig:oasbuilder-system} illustrates the pipeline.


\subsection{Scraping}
\label{subsec:scraping}
The first step is scraping the API documentation webpage, with the goal of segmenting it into continuous sections, each linked to a specific operation through the identification of operation signatures and usage examples.
The application begins by launching a web browser and navigating to the specified URL. To achieve this, it utilizes Selenium,\footnote{\url{https://www.selenium.dev/}} a tool that provides capabilities for controlling a web browser, enabling us to interact with and query the webpage HTML elements.
Since many pages load part of their content dynamically through user interactions, the system uses a small set of rules to detect and click on elements like ``expand all'' or ``example''. 
After the content is loaded, \productname{} identifies operation signatures and API request examples. Specifically, it searches for HTML elements whose text includes cURL commands, HTTP commands, or patterns resembling operation signatures, such as \texttt{<HTTP\_METHOD ENDPOINT>}. 
\productname{} assumes that all instances of a specific operation appear sequentially on the page. Hence, it defines the boundaries of each operation as the section spanning from the first title preceding its initial instance to the first title marking the start of the next operation.

\subsection{Demonstrative OAS Generation}
\label{subsec:openapi-spec-generation}



After segmenting the webpage into operations, as outlined in Section~\ref{subsec:scraping}, the next step (step 2 in Figure~\ref{fig:oasbuilder-system}) involves extracting demonstrative examples and generating an OAS from them.
Since the system has already extracted request examples in the previous stage, it only needs to identify the corresponding request bodies and response examples within the operation boundaries, if available.

After extracting demonstrative examples, the system converts them into OAS format by decomposing the process into multiple parallel LLM calls. We favor LLMs over complex rule-based parsing, as the latter is prone to errors arising from human mistakes in example creation and noise introduced during automated scraping. This multi-stage approach addresses input length limitations while enabling efficient parallel generation.
First, \productname{} generates a partial OAS for each request example, excluding the request body. To do so, each example is standardized into a canonical cURL command to minimize variations. Parallel LLM calls are then employed, each using two diverse in-context examples drawn from real-world scenarios to convert the standardized commands into partial OAS documents. These include essential metadata such as servers, paths, HTTP methods, and request parameters.
Second, the system generates JSON schemas for both request bodies and response examples. Due to the difficulty LLMs face in processing large, deeply nested JSON structures—a well-documented issue~\citep{shorten2024structuredragjsonresponseformatting}—\productname{} divides these structures into smaller fragments based on a predefined line threshold, preserving scope boundaries. Parallel LLM calls are then applied to each fragment, using two in-context examples per call, to generate the corresponding JSON schemas. Prompt examples are provided in Appendix~\ref{subsec:appendix-prompt-generation-examples}.




\subsection{Descriptive OAS Generation}

\label{subsec:enrichment-specrawler}

In parallel with generating an OAS from the demonstrative examples discussed in Section~\ref{subsec:openapi-spec-generation}, \productname{} also generates a second OAS based on the descriptive documentation (see Section~\ref{subsec:openapi-spec-generation} for more details). This stage is labeled as step 3 in Figure~\ref{fig:oasbuilder-system}.

Parsing these documentations using deterministic rule-based algorithms is impractical due to the significant variations in HTML structures across websites~\citep{Yang-2018}. Therefore, leveraging the capabilities of contemporary LLMs offers a more viable solution, as they can generalize over such structural differences effectively.




Therefore, \productname{} first applies a search algorithm to extract the descriptive documentation from the operation's scope identified in the scraping stage.
The algorithm employs various heuristics to identify and filter the appropriate HTML elements, leveraging prior knowledge of the high-level structure of these webpages.
For instance, for a webpage containing a single operation, it identifies the smallest HTML scope that includes both an operation signature or a request example, while maximizing the number of HTML elements that their text is equal to a parameter name from the example. Additionally, usage examples and any HTML elements lacking indicators of relevant information are excluded.
For more details, see Appendix~\ref{subsec:appendix-find-minimal-ancestor} and algorithm~\ref{alg:minimal-ancestor-algorithm}.  

After narrowing the scope to a relatively small number of HTML elements, the LLM input is further reduced by filtering out the HTML attributes (e.g., css styles) as they often do not contain any relevant information. An LLM is then employed using in-context learning to convert the extracted HTML into an OAS. 
We found that LLMs often require exposure to various structures in the examples to correctly apply them in the test case. For example, the model would fail to generate a requestBody component or an enum attribute if they were not provided in the input example.
Therefore, the in-context example was carefully crafted using data from multiple real-world APIs with diverse structures and attributes.
In the case of context window overflow, it retries with an alternative shorter in-context example. 
A prompt example is provided in Appendix~\ref{subsec:appendix-prompt-generation-examples}.
After generating the OAS, the system validates its structure and verify that all the generated parameter names appear in the input to prevent hallucinations. 

Lastly, the generated OAS is merged with the one created in Section~\ref{subsec:openapi-spec-generation} to yield a final comprehensive OAS. In this integration, the description and required attributes from the descriptive documentation are prioritized, while the type and location fields from the demonstrative documentation take precedence. This prioritization is determined based on the reliability of the attribute in each type of documentation.

\subsection{OAS Enhancement}
\label{subsec:enhancement}


After generating an OAS from the documentation, \productname{} enriches missing metadata using AI-based tools based on information within the OAS (step 4 in Figure~\ref{fig:oasbuilder-system}). These tools perform two key functions: (1) extracting parameter metadata from grounded parameter descriptions and (2) generating missing metadata based on its surrounding context. This enriched metadata, including descriptions, enums, and defaults, is essential for accurately documenting API behavior and supports various downstream tasks, such as conversational agents, slot filling~\citep{vaziri2017generating}, test generation~\citep{kim2022}, and API sequencing.


\paragraph{Metadata extraction from parameter descriptions:} 
Parameter descriptions often include metadata like enum, default, and format, which can be explicitly defined in the OAS. LLMs are well-suited for extracting this metadata. To improve extraction, we designed prompts using in-context examples that handle both explicit (e.g., "the default value is 10") and implicit metadata (e.g., "the option is disabled by default"). 
To avoid hallucinations, the extracted values are verified to match the descriptions. To minimize LLM calls, \productname{} used two strategies: First, a keyword-based filtering mechanism that triggers LLM calls only when relevant terms like "default" or "not provided" are present, saving over 90\% of calls for many metadata fields. Second, the prompts were designed to process multiple descriptions at once, further reducing LLM calls.


\paragraph{Metadata Generation Using OAS Structure:} \productname{} addresses missing method and parameter descriptions, as well as parameter examples, in an OAS by utilizing LLM-based prompts to generate the missing metadata. Relevant context from the OAS is extracted and provided as input to these prompts. For example, the context for generating parameter descriptions and examples are the parameter name, the method name and description and the parameter description for the example generation. For method descriptions, it incorporates the method name, endpoint path, and operation ID. 
Examples of the generated metadata is provided in Appendix~\ref{subsec:appendix-generated-examples}.

%% file: oasbuilder/chapters/4-results.tex
This section presents the results of a series of experiments conducted to evaluate the performance of \productname{}.
We utilized several well-known open-source LLMs with in-context prompting capabilities, including llama-3-70b-instruct~\citep{touvron2023llama}, codellama-34b-instruct~\citep{rozière2024code}, mistral-7b-instruct~\citep{jiang2023mistral}, mixtral-8x7b-instruct~\citep{jiang2024mixtral}, and 
granite-20b-code-instruct~\citep{mishra2024granitecodemodelsfamily}. 
The prompts were not fine-tuned for any specific model. Baselines were not included, as previous studies neither evaluated on a public benchmark nor provided their code or reproduction details.


\subsection{Syntactic Evaluation}
\label{subsec:results-base-oas-generation}
In this section, we analyze the syntactic properties of OAS documents generated by \productname{} using various LLMs on a corpus of 50 diverse documentation webpages covering 189 operations (see Appendix~\ref{subsec:appendix-50-urls-base-oas-generation} for details).

The evaluation focuses on three metrics: (1) the proportion of outputs that are valid JSONs; (2) the proportion that qualify as valid OAS documents; and (3) the average number of errors in valid JSONs. While the latter two metrics are related—errors occur only in invalid OAS documents—quantifying the errors provides insight into the degree of syntactic deviation, helping to estimate the effort required for correction.
We computed the two metrics with jsonschema library.\footnote{\url{https://github.com/python-jsonschema/jsonschema}}

Table~\ref{tab:Generation_Results} presents the results of the syntactic analysis. Notably, granite-code and codellama emerge as the top-performing models. This likely reflects the prevalence of JSON-related tasks in code-oriented benchmarks. While codellama achieved the highest proportion of valid OAS, its error rate ranked third among the models. In contrast, granite-code produced the second-highest rate of valid OAS while exhibiting the lowest error incidence. The remaining models generally succeeded in generating valid JSON but showed considerably lower and more variable rates of valid OAS generation. These findings suggest that, even when decomposed into subtasks, OAS generation remains a nontrivial challenge for LLMs.

To evaluate scalability, we collected a larger dataset of 291 API documentation URLs and repeated the experiment using the granite-code model. Results showed that 100\% of the outputs were valid JSON, 89\% were valid OAS, and the average number of errors per OAS was 0.17. Furthermore, 86\% of the OAS documents contained at least one operation and one parameter.

Lastly, we attempted to generate OAS documents using GPT-4-128K~\citep{openai2024gpt4technicalreport} directly from the original HTML, without using \productname{}. We found that the model was able to generate a valid OAS only for $25\%$ of the webpages. This outcome is not unexpected, as many of these webpages contain much more than 128K tokens, the model's context window limit. 

\begin{table}[t]
\begin{center}
\begin{small}
\begin{sc}
\begin{tabular}{@{}llll@{}}
\toprule
                     & Valid JSON          & Valid OAS       & Errors   \\ \midrule
codellama    &  $.99$     &  \boldsymbol{$.89$}  &  $.59$ \\
granite-code      &  \boldsymbol{$1$} &  $.73$     &  \boldsymbol{$.48$}  \\
llama-3    &  \boldsymbol{$1$}  &  $.29$    &  $.78$  \\
mistral      &  \boldsymbol{$1$}     &  $.4$  &  $.54$ \\
mixtral      &  $.92$   &  $.66$ &  $.64$ \\ \bottomrule
\end{tabular}
\caption{Syntactic evaluation results for OAS generation by different LLMs on 50 web pages covering 189 operations. Metrics include the ratios of valid JSONs and OAS documents and the average errors in valid JSONs.
}
\label{tab:Generation_Results}

\end{sc}
\end{small}
\end{center}
\end{table}

\subsection{Semantic Evaluation}
\label{subsec:end-to-end-results}
\input{oasbuilder/tables/end_to_end_table}

In this section, we assess the overall capabilities of \productname{} to generate rich and complete OAS given various API documentation webpages.
To that end, we employed a manually labeled dataset comprising of 108 operations containing thousands of parameters and properties from different API documentation websites.
We conducted experiments to compare the enhanced OAS documents generated by \productname{} with the ground truth OAS, using different LLMs. 
In all experiments we used the in-context learning approach with the same prompt and in-context examples across models.

Table~\ref{tab:e2e} presents the end-to-end results.
First, the parameter precision for all models was relatively high, suggesting that hallucinations were uncommon. The recall for request parameters was also high, particularly for granite-code and code-llama, with values of 0.86 and 0.85, respectively. Additionally, the description similarity and the F1 scores for the required, default, and enum attributes were relatively high.
These findings indicate that, although the generated OAS documents are not perfect, they capture most of the relevant information on the request side, significantly reducing the user's manual annotation effort.
Since many request parameters and their attributes such as default, enum, and description are found exclusively in the descriptive documentation, we can conclude that the information from the descriptive documentation were successfully integrated in the final OAS.

Lastly, the recall for response generation was lower, likely due to the highly nested and lengthy structure of many responses, as well as the frequent lack of descriptive documentation for response properties. Overall, the LLMs demonstrated competitive performance, with no single model showing clear dominance, though mistral and mixtral performed slightly below the others.


%% file: oasbuilder/tables/end_to_end_table.tex
\begin{table*}[t]
\begin{center}
\begin{small}
\begin{sc}
\resizebox{\linewidth}{!}{%
\begin{tabular}{@{}lllllllllllll@{}}
\toprule
          & \multicolumn{7}{c}{Request}                                                                                           & \multicolumn{5}{c}{Response}                                     \\ \midrule
model     & P            & R            & F1           & Desc.         & Req.     & Def.      & Enum             &           & P            & R            & F1           & Desc.      \\ \midrule
codellama & .95          & .85 & .90 & .89          & \textbf{.88} & .88          & .69                    &           & .93          & .56          & .70          & .68       \\ 
granite-code   & \textbf{.96} & \textbf{.86}          & \textbf{.91}          & .90          & \textbf{.88} & .84          & .76           &           & .92          & .54          & .68          & .79       \\
llama-3   & \textbf{.96} & .78          & .86          & \textbf{.91} & .87          & .92          & \textbf{.81}                   & \textbf{} & \textbf{.97} & \textbf{.62} & \textbf{.75} & .72   \\
mistral   & .94          & .67          & .78          & .75          & .85          & .56          & .50           &           & .90          & .57          & .69          & .64 \\
mixtral     & .95          & .55          & .70          & .87          & .80          & \textbf{.93} & .62                    &           & .90          & .52          & .65          & \textbf{.84}  \\ \bottomrule
\end{tabular}
}
\caption{End-to-end results of \productname{} for different LLMs on 108 different operations containing thousands of parameters and properties. We report the precision (P), recall (R), F1-score (F1) of the parameters, and the cosine similarity of the descriptions, as well as the F1-score of required (Req.), default (Def.) and enum attributes. All results are averaged across all parameters and were based on the valid OASs for each model. 
}
\label{tab:e2e}
\end{sc}
\end{small}
\end{center}
\end{table*}

%% file: oasbuilder/chapters/5-related-work.tex
Varied methods have been adopted to generate OAS documents for Rest APIs. SpyREST~\citep{sohan2015spyrest} employs an HTTP proxy server to intercept HTTP traffic to generate API documentation. Respector~\citep{huang2024generating} employs static and symbolic program analysis to automatically generate OAS for REST APIs from their source code.

Similar to our approach, several studies have investigated converting parts of API documentation webpages into OAS. Auto\-REST \citep{cao2017automated} and captures part of the information presented in API documentation webpages and converts it into an OAS by a set of fixed rules. D2Spec~\citep{Yang-2018} aims to extract base URLs, path templates, and HTTP method types, using rule-based web crawling techniques and classic machine learning to identify potential API call patterns in URLs.
\citet{bahrami2020automated, bahrami2020deep} combines rule-based and machine-learning algorithms to generate OAS from API documentation. They also develop a deep model to pinpoint fine-grained mapping of extracted API attributes to OAS objects. 
Most similar to our work, \citet{androvcec2023using} used GPT-3 to automatically generate OAS from a preprocessed HTML file describing an API documentation. \productname{} distinguishes itself from their methodology by (1) dividing the generation task into multiple parts, and (2) extracting relevant information from webpages, thus accommodating long webpages that exceeds the context size of LLMs while breaking the task into more manageable subtasks for LLMs..

%% file: oasbuilder/chapters/6-conclusions.tex
This paper presents \productname{}, a novel multi-stage system designed to automatically generate and enhance OAS from online API documentation. By integrating rule-based algorithms with generative LLMs, \productname{} addresses existing limitations in previous solutions. Our experiments demonstrate that \productname{} is robust and capable of generalizing across hundreds of API documentation websites. Furthermore, a detailed evaluation reveals that the generated OAS captures most of the information from the documentation, significantly reducing the manual effort required of technical experts. The AI-based enhancement tools, combined with the UI platform, offer developers an end-to-end process yielding in a high-quality final result.

%% file: oasbuilder/chapters/7-appendix.tex
\subsection{Example of an API documentation webpage}
\label{subsec:appendix-api-documentation-webpage}
\begin{figure*}
    \centering
     \includegraphics[width=1.\linewidth]{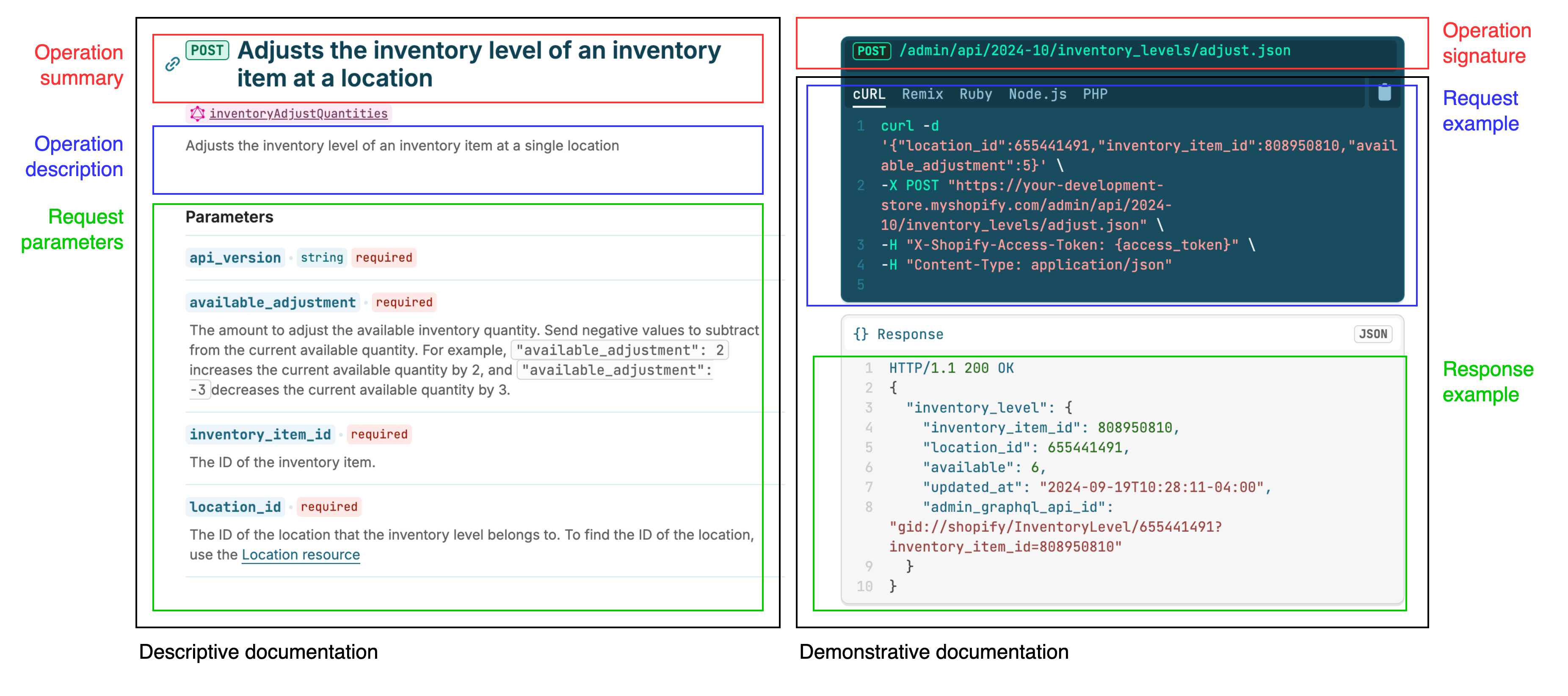}
    \caption{\label{fig:shopify-documentation-webpage-example}Example of an API documentation webpage taken from Shopify API in 25.11.2024 (taken from \url{https://shopify.dev/docs/api/in-rest/2024-10/resources/inventorylevel}.}
\end{figure*}

Figure~\ref{fig:shopify-documentation-webpage-example} shows an example of a real-world API documentation webpage taken from Shopify API website.\footnote{\url{https://shopify.dev/docs/api/admin-rest/2024-10/resources/inventorylevel}}

\subsection{Descriptive Documentation Retrieval Algorithm}
\label{subsec:appendix-find-minimal-ancestor}
To find the descriptive documentation, we looked for an HTML scope in the webpage which encompasses this information, we call this scope ``minimal ancestor''. To that end, we employed two distinct approaches. In scenarios where a webpage incorporates multiple API calls, we defined the scope as the highest ancestor of the request example HTML element that does not encompass other requests~\footnote{If the minimal ancestor of the subsequent request is not consecutive, it is defined as a sequence of elements ending in the ancestor of the next request}. In Figure~\ref{fig:webpage-example}, this scope should encompass both reference-based and example-style sections. Conversely, when dealing with a webpage containing a single API call, we conducted a search for \textit{leaf elements.}\footnote{HTML elements lacking children} likely associated with parameters in the reference-based documentation based on their text, such as parameters from the request or response, and parameter header templates. These elements could be situated, for instance, in the ``Parameters Description'' section as illustrated in Figure~\ref{fig:webpage-example}. Subsequently, we iterated through the ancestors of each identified element, starting from the immediate parent and moving upwards, in search of the first ancestor containing a matching URL endpoint corresponding to the provided API URL. Since this is often found preceding the HTTP method (e.g. ``GET /info/{id}''), we denote it as ``HTTP Method'' and ``Endpoint/URL'' in Figure~\ref{fig:webpage-example}.

After retrieving these minimal ancestors, we rank them according to two criteria: 1) the number of parameters from the request or response found as leaf elements in the ancestor, and 2)
whether the HTTP method type of the URL was found as a leaf element. 
Following this ranking, we filter out HTML elements that are ancestors of other candidates. 
Lastly, if we still have multiple candidates sharing the same rank, we randomly sample one of them, although we did not encounter such cases in our experiments.

The minimal ancestor is then preprocessed to remove noise and tailor it to the constrained context size of the LLM. This involves filtering out its children that are less likely to contain relevant information for augmenting the base OAS. Specifically, we search for parameter names extracted from the API request/response example and syntactic hints such as the structure of an HTML parameters table. Additionally, we exclude the request and response examples at this stage, as they have already been utilized in generating the base OAS. Finally, all HTML attributes are removed, as they are deemed less likely to contain relevant information.

For a formalized presentation of the flow, see Algorithm~\ref{alg:minimal-ancestor-algorithm}

\begin{algorithm*}
\caption{Generate Descriptive Documentation}
\hspace*{\algorithmicindent} \textbf{Input:}  Example of API Request, and API HTML Documentation \\
\hspace*{\algorithmicindent} \textbf{Output:} HTML Element for Enrichment
\begin{algorithmic}[1]

\Function{FindMinimalHTMLElement}{API\_Spec, API\_Doc, M}
    \State 1. Extract parameter names from given API request.
    \State 2. Find elements in documentation that their texts match a parameter name or a parameter header.
    \State 3. For each candidate find first HTML elements which meets one of the following criteria:
    \State \quad a. Contains an endpoint HTML element matching the API URL from the request.
    \State \quad b. Contains multiple HTML elements of the same parameter name from the API specification.
    \State \quad iv. Select the HTML element from the candidates by ranking according to the following criteria by the following order:
    \State \quad \quad 1. Number of parameter names from the API specification found in its context by exact matching*.
    \State \quad \quad 2. Whether an endpoint matching the URL was found.
    \State \quad \quad 3. Whether the extracted HTTP method type was found by exact matching.
    \State \quad \quad 4. Whether they contain a "table" HTML element.
    \State \quad \quad 5. Minimality of scope (i.e. filtering out parents of candidates).
    \State \textbf{Preprocessing the minimal HTML element:}
    \State \quad i. Iterate over the minimal HTML element children and filter according to the following criteria:
    \State \quad \quad 1. Whether the child is a "table" HTML element.
    \State \quad \quad 2. Whether the child is preceded by a parameter header HTML element.
    \State \quad \quad 3. Whether the child contains any extract parameter name by exact matching.
    \State \quad \quad 4. Whether the child contains the phrases "required" or "optional" by exact matching.
    \State \quad ii. Remove all the attributes of the HTML elements.
    \State \textbf{Generate Structured data from minimal HTML element:}
    \State \quad i. Apply M to generate a description of the API and a table where each row represents relevant metadata about a parameter found in the minimal HTML element's content. This can be achieved by techniques such as In-Context Learning, or by training a language model on a manually-labeled dataset.
    \State \textbf{Integrate Generated Data into API Specification.}
\EndFunction
\end{algorithmic}
\label{alg:minimal-ancestor-algorithm}
\end{algorithm*}

\subsection{Prompt Generation Examples}
\label{subsec:appendix-prompt-generation-examples}

\begin{figure*}
     \includegraphics[width=2\columnwidth]{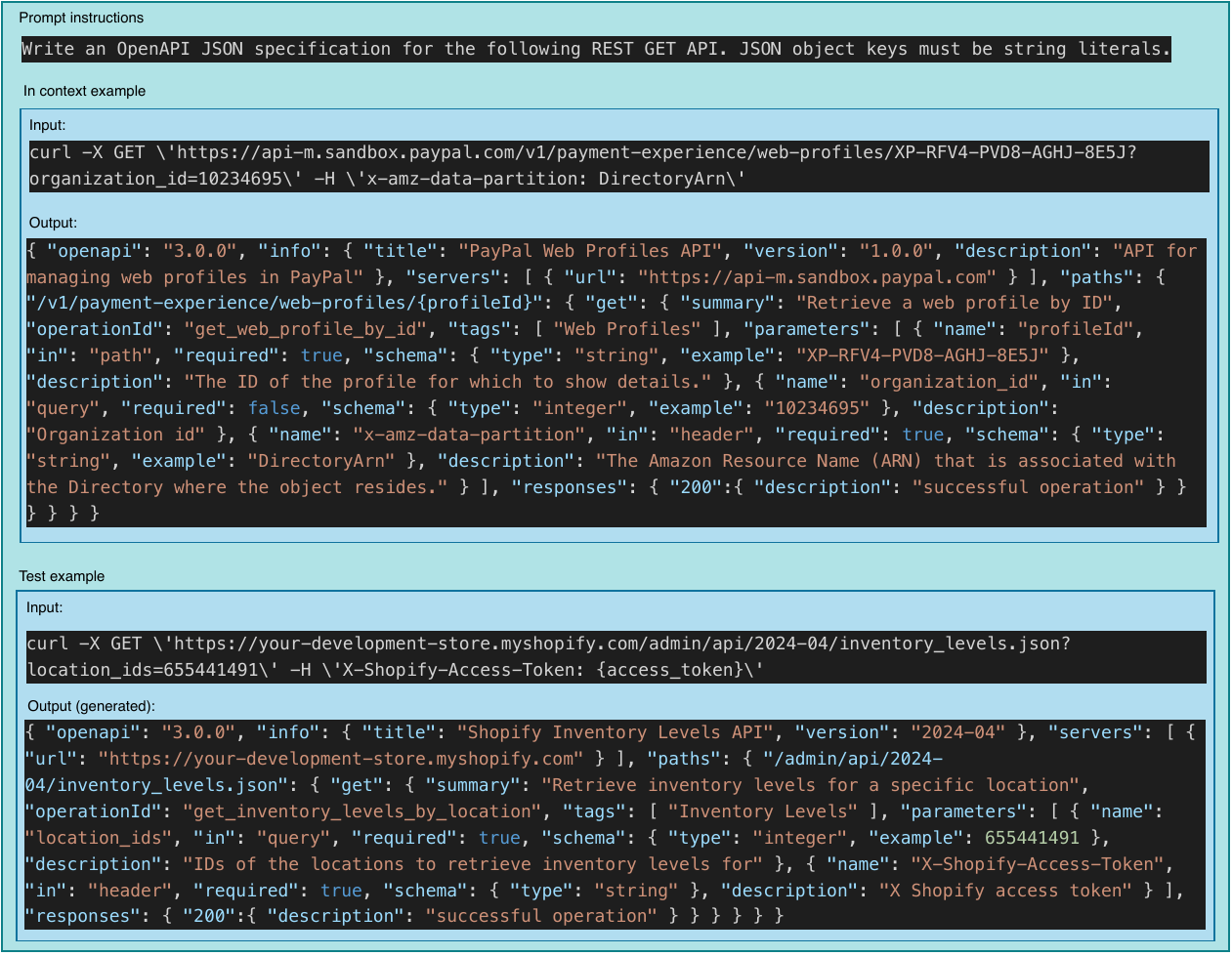}
    \caption{\label{fig:oas-prompt} Example prompt for generating an OAS from a cURL command. The prompt includes two in-context examples (only one is shown here for brevity). Information from the cURL command is extracted to create an OAS featuring a single operation, complete with a \texttt{title}, \texttt{version}, \texttt{servers}, \texttt{paths}, \texttt{operationId}, \texttt{tags}, and detailed \texttt{parameters}, including their \texttt{type}s, \texttt{description}s, and \texttt{example}s.
    } 
\end{figure*}

\begin{figure*}
     \includegraphics[width=2\columnwidth]{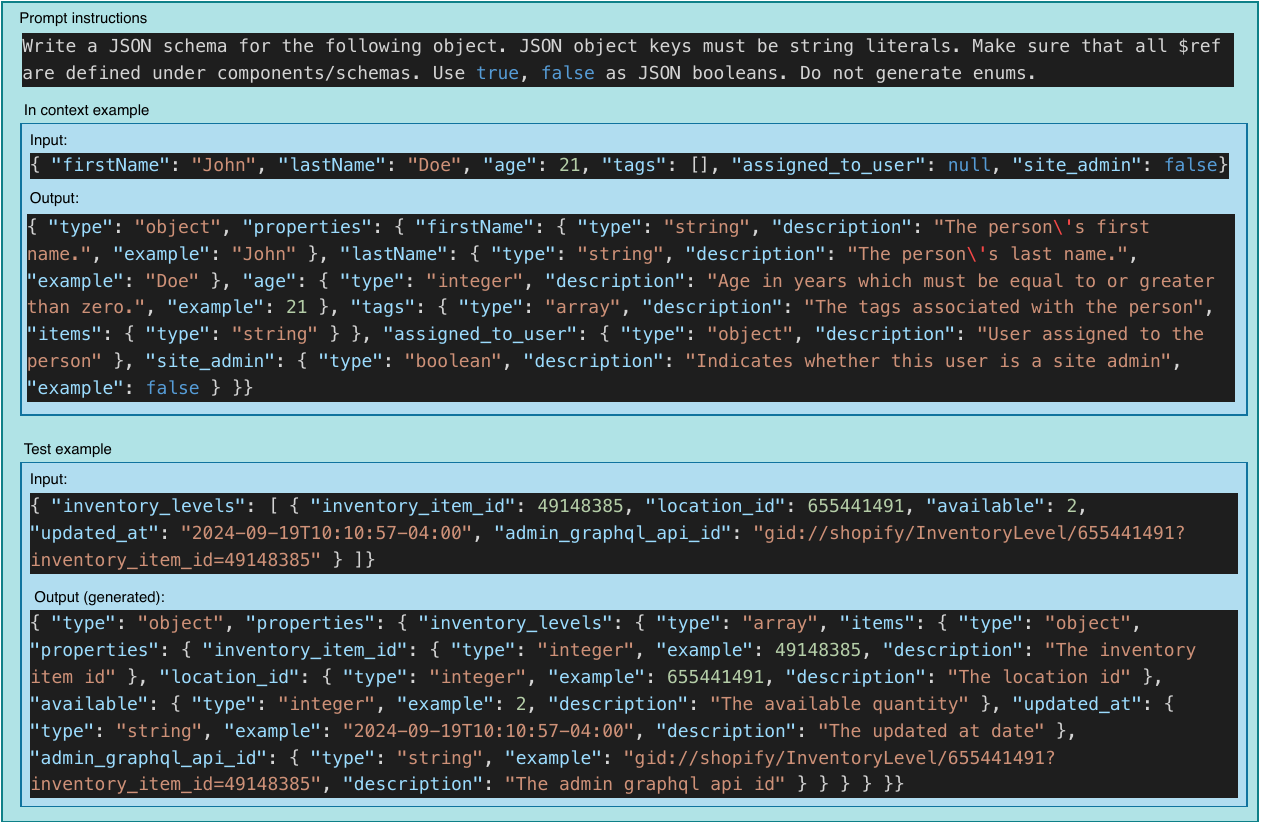}
    \caption{\label{fig:jschema-prompt} Prompt example to generate a JSON schema from a given JSON object or array. This prompt is used to generate both the \texttt{requestBody} and the \texttt{response}s which are later set in the corresponding OAS.
    } 
\end{figure*}

\begin{figure*}
     \includegraphics[width=2\columnwidth]{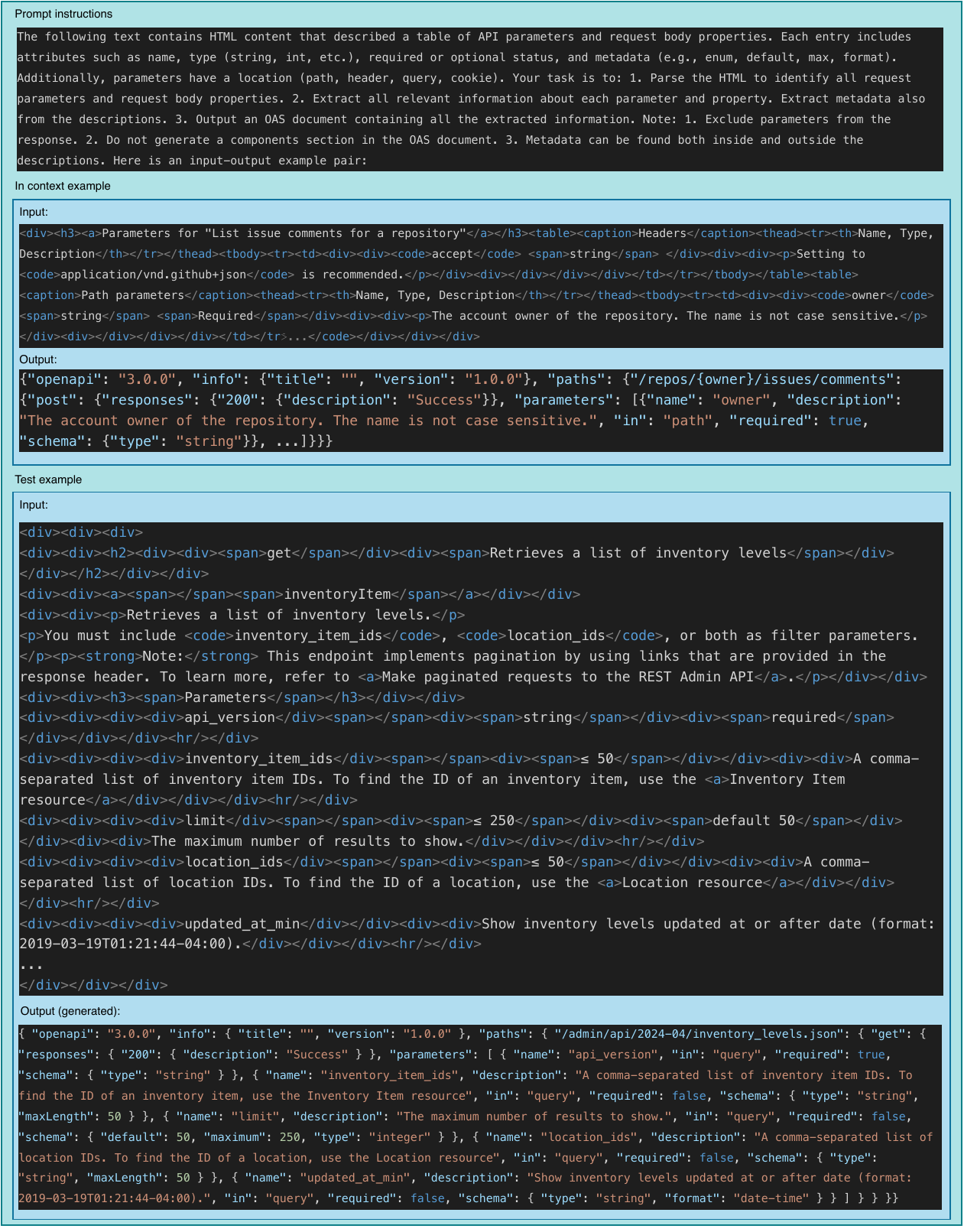}
    \caption{\label{fig:request-enrichment-prompt} An example of a prompt used for generating an OAS based on the descriptive documentation found in the API documentation webpage. The model extract the relevant information from the HTML elements, and sets the fields' \texttt{description}, \texttt{type}, \texttt{required}, \texttt{enum}, and \texttt{format} metadata properties.
    } 
\end{figure*}

In order to generate the OAS, we applied in-context learning where the inputs are preprocessed content found in the API documentation webpage, and the outputs are components from the OAS or partial OAS containing the input data. The in-context examples were chosen from real-world APIs (e.g., github API) while we tried to balance between the length and the diversity of the examples.
In Figures~\ref{fig:request-enrichment-prompt},~\ref{fig:oas-prompt},~\ref{fig:jschema-prompt} we provide examples of the prompts we used for this purpose. Due to space limitations, we have not included all the in-context examples, but we would be happy to share them upon request.

\subsection{URLs for Base OAS Generation}
\label{subsec:appendix-50-urls-base-oas-generation}
\begin{itemize}
    \item \url{https://docs.sendgrid.com/api-reference/contacts/add-or-update-a-contact}
    \item \url{https://developer.servicenow.com/dev.do#!/reference/api/sandiego/rest/c_TableAPI}
    \item \url{https://dev.fitbit.com/build/reference/web-api/activity/get-activity-log-list/}
    \item \url{https://docs.adyen.com/api-explorer/Checkout/70/post/payments}
    \item \url{https://openweathermap.org/api/one-call-3}
    \item \url{https://developer.cisco.com/meraki/api-v1/get-device-camera-custom-analytics/}
    \item \url{https://developer.paypal.com/docs/api/payment-experience/v1/#web-profile_create}
    \item \url{https://stripe.com/docs/api}
    \item \url{https://developer.webex.com/docs/api/v1/meeting-transcripts/download-a-meeting-transcript}
    \item \url{https://developer.okta.com/docs/api/openapi/okta-management/management/tag/ApiServiceIntegrations/#tag/ApiServiceIntegrations/operation/activateApiServiceIntegrationInstanceSecret}
    \item \url{https://developer.okta.com/docs/api/openapi/okta-management/management/tag/ApplicationGroups/#tag/ApplicationGroups/operation/assignGroupToApplication}
    \item \url{https://learn.microsoft.com/en-us/linkedin/shared/integrations/communications/invitations?context=linkedin%2Fcompliance%2Fcontext&view=li-lms-unversioned&preserve-view=true}
    \item \url{https://www.aha.io/api/resources/ideas/create_an_idea}
    \item \url{https://www.reddit.com/dev/api}
    \item \url{https://cloud.ibm.com/apidocs/speech-to-text}
    \item \url{https://apidocs.orderdesk.com/?shell#create-an-order}
    \item \url{https://developer.atlassian.com/cloud/trello/rest/api-group-actions/#api-actions-idaction-reactions-post}
    \item \url{https://docs.github.com/en/rest/issues/comments?apiVersion=2022-11-28#create-an-issue-comment}
    \item \url{https://docs.github.com/en/rest/actions/workflow-runs?apiVersion=2022-11-28#re-run-a-job-from-a-workflow-run--code-samples}
    \item \url{https://developers.facebook.com/docs/whatsapp/business-management-api/guides/migrate-phone-to-different-waba}
    \item \url{https://docs.github.com/en/rest/issues/issues?apiVersion=2022-11-28}
    \item \url{https://community.workday.com/sites/default/files/file-hosting/restapi/index.html}
    \item \url{https://community.workday.com/sites/default/files/file-hosting/restapi/index.html}
    \item \url{https://community.workday.com/sites/default/files/file-hosting/restapi/index.html#budgets/v1/post-/runBudgetCheck}
    \item \url{https://docs.sendgrid.com/api-reference/contacts/delete-contacts}
    \item \url{https://docs.sendgrid.com/api-reference/custom-fields/update-custom-field-definition}
    \item \url{https://docs.sendgrid.com/api-reference/custom-fields/create-custom-field-definition}
    \item \url{https://shopify.dev/docs/api/admin-rest/2023-04/resources/asset#put-themes-theme-id-assets}
    \item \url{https://shopify.dev/docs/api/admin-rest/2023-04/resources/product#put-products-product-id}
    \item \url{https://docs.mapbox.com/api/search/geocoding/}
    \item \url{https://developers.facebook.com/docs/whatsapp/business-management-api/message-templates}
    \item \url{https://wit.ai/docs/http/20230215/#post__utterances_link}
    \item \url{https://www.twilio.com/docs/sms/api/deactivations-resource}
    \item \url{https://www.twilio.com/docs/sms/api/media-resource}
    \item \url{https://www.twilio.com/docs/sms/api/message-resource#read-multiple-message-resources}
    \item \url{https://airtable.com/developers/web/api/delete-multiple-records}
    \item \url{https://airtable.com/developers/web/api/update-record}
    \item \url{https://airtable.com/developers/web/api/refresh-a-webhook}
    \item \url{https://developer.cisco.com/meraki/api-v1/blink-device-leds/}
    \item \url{https://developer.cisco.com/meraki/api-v1/get-network-events/}
    \item \url{https://developer.cisco.com/meraki/api-v1/get-organization-summary-top-appliances-by-utilization/}
    \item \url{https://docs.github.com/en/free-pro-team@latest/rest/billing/billing?apiVersion=2022-11-28#get-github-actions-billing-for-an-organization}
    \item \url{https://docs.github.com/en/rest/issues/comments?apiVersion=2022-11-28}
    \item \url{https://docs.github.com/en/rest/interactions/user?apiVersion=2022-11-28}
    \item \url{https://docs.github.com/en/rest/search/search?apiVersion=2022-11-28}
    \item \url{https://learn.microsoft.com/en-us/linkedin/shared/api-guide/concepts/pagination?context=linkedin%2Fconsumer%2Fcontext}
    \item \url{https://dev.fitbit.com/build/reference/web-api/sleep/delete-sleep-log/}
    \item \url{https://dev.fitbit.com/build/reference/web-api/body/create-bodyfat-log/}
    \item \url{https://dev.fitbit.com/build/reference/web-api/friends/get-friends-leaderboard/}
\end{itemize}

\subsection{Examples of Generated Descriptions and Examples}
\label{subsec:appendix-generated-examples}
Figure~\ref{fig:example-description} and Figure~\ref{fig:example-example} are respectively examples of generated descriptions and examples from the enhancements described in Section~\ref{subsec:enhancement}. 

\begin{figure*}
\centering
\begin{lstlisting}
{
    "200": {
        "content": {
            "application/json": {
                "schema": {
                    "properties": {
                        @ "dateLastActivity": {
                            "type": "string",
                            "description": "The date the activity was last updated."
                        }, @
                        "dateLastView": {
                            "type": "string",
                            "description": "The last time the user viewed the board."
                        },
                        @ "idTags": {
                            "type": "string",
                            "description": "A comma-separated list of tag IDs. Only actions within these tags will be returned."
                        } @
                    }
                }
            }
        }
    }
}
\end{lstlisting}
\caption{\label{fig:example-description}Example of enhancement for generating descriptions. Added lines are highlighted in green. Original documentation page for OAS is \url{https://developer.atlassian.com/cloud/trello/rest/api-group-actions/\#api-actions-idaction-reactions-post}.}
\end{figure*}

\begin{figure*}
\centering
\begin{lstlisting}
{
    "parameters": [
        {
            "name": "owner",
            "in": "path",
            "required": true,
            "schema": {
                "type": "string",
                @ "example": "octocat",
                "x-ibm-examples": [
                    "hubot",
                    "other_user"
                ] @
            },
            @ "description": "The account owner of the repository. The name is not case sensitive.",
            "x-ibm-grounded-description": true @
        },
        {
            "name": "repo",
            "in": "path",
            "required": true,
            "schema": {
                "type": "string",
                @ "example": "octocat/Hello-World",
                "x-ibm-examples": [
                    "octocat/Spoon-Knife",
                    "octocat/hello-world"
                ] @
            },
            @ "description": "The name of the repository without the \".git\" extension. The name is not case sensitive.",
            "x-ibm-grounded-description": true @
        }
    ]
}
\end{lstlisting}
\caption{\label{fig:example-example}Example of enhancement for generating examples. Added lines are highlighted in green. Original documentation page for OAS is \url{https://docs.github.com/en/rest/issues/comments?apiVersion=2022-11-28\#create-an-issue-comment}.}
\end{figure*}

\subsection{Most Popular URLs by Postman}
\label{subsec:appendix-most-popular-urls}
To further establish the claim that most real-world APIs do not publish API specification. We manually checked whether the most popular APIs according to Postman\footnote{\url{https://www.postman.com}} published an API specification in their API documentation webpages. We found that only five out of the fourteen contained OAS. The full findings are detailed in Table~\ref{tab:popular_apis} 

\begin{table*}[t]
\centering
\begin{tabular}{|l|p{0.45\textwidth}|c|}
\hline
\textbf{Site}       & \textbf{API Documentation URL}                                     & \textbf{Contains OAS} \\ \hline
Salesforce          & \url{https://developer.salesforce.com/docs/apis\#browse}            & No                    \\ \hline
Microsoft Graph     & \url{https://learn.microsoft.com/en-us/graph/overview}              & No                    \\ \hline
Slack               & \url{https://api.slack.com/docs/apps}                               & No                    \\ \hline
PayPal              & \url{https://developer.paypal.com/api/rest/}                        & No                    \\ \hline
Zoho CRM            & \url{https://www.zoho.com/crm/developer/docs/api/v7/modules-api.html} & No                    \\ \hline
Cisco Meraki        & \url{https://developer.cisco.com/meraki/}                            & Yes                   \\ \hline
Pipedrive API       & \url{https://developers.pipedrive.com/docs/api/v1}                  & Yes                   \\ \hline
Amplitude           & \url{https://amplitude.com/docs/apis/analytics}                     & No                    \\ \hline
BookingAPI          & \url{https://developers.booking.com/demand/docs}                    & Yes                   \\ \hline
Amadeus             & \url{https://developers.amadeus.com/self-service}                   & Yes                   \\ \hline
Symbl               & \url{https://docs.symbl.ai/reference}                               & No                    \\ \hline
Hyperledger Besu    & \url{https://besu.hyperledger.org/stable/public-networks/reference/api} & No                    \\ \hline
PingOne             & \url{https://apidocs.pingidentity.com/pingone/platform/v1/api/}     & No                    \\ \hline
Lob                 & \url{https://docs.lob.com/}                                         & Yes                   \\ \hline
\end{tabular}
\caption{Comparison of the most popular APIs on Postman for 2023, indicating whether they publicly publish their OAS (based on \url{https://www.postman.com/explore/most-popular-apis-this-year}).}
\label{tab:popular_apis}
\end{table*}